\documentclass[aps,prb,twocolumn,noshowpacs,superscriptaddress]{revtex4-1}
\usepackage{amsmath}
\usepackage{graphicx}
\usepackage{amsbsy}
\usepackage{amssymb}
\usepackage{amsmath}
\usepackage{siunitx}
\usepackage{epsfig}
\usepackage{latexsym}
\usepackage{wasysym}
\usepackage{pifont}
\usepackage{mathrsfs}
\usepackage{natbib}
\usepackage{float}
\newfloat{widefig}{thp}{lop}
\usepackage{comment}
\usepackage{verbatim}
\usepackage{multirow,array}
\usepackage{hyperref}
\usepackage{braket}

\renewcommand{\vec}[1]{\mathbf{#1}}
\usepackage{xcolor}

\begin{document}

\title{Effect of local chemistry and structure on thermal transport in doped GaAs}

\author{Ashis Kundu}
\email[]{ashis.kundu@tuwien.ac.at}
\affiliation{Institute of Materials Chemistry, TU Wien, A-1060 Vienna, Austria}
\affiliation{Institute for Advanced Study, Shenzhen University, Nanhai Avenue 3688, Shenzhen 518060,
People's Republic of China}
\author{Fabian Otte}
\author{Jes\'us Carrete}
\affiliation{Institute of Materials Chemistry, TU Wien, A-1060 Vienna, Austria}
\author{Paul Erhart}
\affiliation{Chalmers University of Technology, Department of Physics, Gothenburg, Sweden}
\author{Wu Li}
\affiliation{Institute for Advanced Study, Shenzhen University, Nanhai Avenue 3688, Shenzhen 518060,
People's Republic of China}
\author{Natalio Mingo}
\affiliation{LITEN, CEA-Grenoble, 38054 Grenoble Cedex 9, France.}
\author{Georg K. H. Madsen}
\affiliation{Institute of Materials Chemistry, TU Wien, A-1060 Vienna, Austria}

\date{\today}

\begin{abstract} 
Using a first-principles approach, we analyze the impact of \textit{DX} centers formed by S, Se, and Te dopant atoms on the thermal conductivity of GaAs. Our results are in good agreement with experiments and unveil the physics behind the drastically different effect of each kind of defect. We establish a causal chain linking the electronic structure of the dopants to the thermal conductivity of the bulk solid, a macroscopic transport coefficient. Specifically,  the presence of lone pairs leads to the formation of structurally asymmetric \textit{DX} centers that  cause resonant scattering of incident phonons. The effect of such resonances is magnified when they affect the part of the spectrum most relevant for thermal transport. We show that these resonances are associated with localized vibrational modes in the perturbed phonon spectrum. Finally, we illustrate the connection between flat adjacent minima in the energy landscape and resonant phonon scattering through detailed analyses of the energy landscape of the defective structures.
\end{abstract}
\maketitle


\section{INTRODUCTION}
The properties of charge carriers in semiconductors determine their suitability for specific electronic and optoelectronic applications.\cite{Sze_Book06,Brennan_Book99} Defects and impurities provide a path for their control but can also introduce unwanted behavior by scattering or capturing electrons or holes, thereby limiting the carrier concentrations and mobilities.\cite{Queisser_Science98} Furthermore, defects in semiconducting materials can also have a significant effect on thermal transport, which is typically dominated by phonons.\cite{Callaway_PRB60,Katre_PRL17,Katre_PRM18,Polanco_PRB18}  

GaAs is one of the most prominent members of the family of group III-V semiconductors. Introducing electron-donating defects by substituting group-VI atoms, i.e., S, Se or Te, on the As position seems like an obvious route to obtain $n$-type conductivity. However, VI-doped GaAs (GaAs:VI) exhibits deep donor states, which are attributed to the formation of so-called \textit{DX} centers. \textit{DX} centers can be thought of as a defect complex where the defect, $D$, is accompanied by a unknown lattice distortion, $X$, which acts as an acceptor. Understanding and controlling the formation of the \textit{DX} centers has been pursued for decades.\cite{Mooney_JAP90} 

The experimental data on thermal conductivity for GaAs:VI is intriguing.\cite{CarlsonJAP65,VuillermozPSSA77} At \SI{40}{K}, the experimental lattice thermal conductivity is approximately \SI{300}{W.K^{-1}.m^{-1}} lower in GaAs:S compared to Se- and Te-doped GaAs, despite the carrier concentration being almost an order of magnitude lower in the GaAs:S sample ($6\times 10^{16}$~cm$^{-3}$ compared to around $10^{18}$~cm$^{-3}$).\cite{CarlsonJAP65,VuillermozPSSA77}
One possible explanation could be that the carrier concentration obtained from Hall coefficient measurements does not correspond to the actual defect concentrations. This would be the case if acceptor \textit{DX} centers compensate the electron doping substitutional defects. Another possible explanation could be the presence of exceptionally strong phonon scatterers. Some of us have recently investigated the effect of doping on the thermal conductivity ($\kappa$) of 3C-SiC.\cite{Katre_PRL17} Far from being a relatively universal function of defect concentration, the reduction of $\kappa$ due to doping was found to have an intricate dependence on the chemical nature of the dopant. In particular, boron doping in 3C-SiC leads to resonant scattering at low frequencies that drastically hinders thermal transport.\cite{Katre_PRL17} We have proposed degenerate adjacent minima in the energy landscape for the substitution as a possible explanation of the low-frequency resonances.\cite{Dongre_JMCC18} In this context the several proposed structures with similar energies\cite{ParkPRB96,DuPRB05,MaPRB13} for the \textit{DX} centers in GaAs:VI  are interesting.
 
The lower thermal conductivity of the GaAs:S sample could well be the signature of a resonance breaking the usual Rayleigh behavior of the acoustic-mode scattering. This interpretation would give additional support to the proposed connection between the minima structure of the potential energy surface and the presence of resonances in the scattering rates, thereby providing a simple and intuitive way of identifying systems which will exhibit resonant phonon scattering.
   
In the present paper, we investigate whether the simple picture of competing minima holds for the complex structure of GaAs:VI.
We explore the influence of \textit{DX} centers formed through S, Se and Te doping at the As site in GaAs on its lattice thermal conductivity. The role of lone pair electrons in the formation of the \textit{DX} centers is illustrated using the electron localization function (ELF).\cite{ELF} We compare our calculated $\kappa$ with available experimental results\cite{CarlsonJAP65,VuillermozPSSA77} and show how the formation of those centers and the associated lattice distortion causes a significant reduction of thermal conductivity in GaAs. The reduction is related to the presence of resonances in the phonon scattering rates at low frequencies. We calculate the perturbed Green's functions of the systems and show how the resonances represent localized vibrational modes directly connected to the aforementioned distortion. The energy landscape of the \textit{DX} centers is explored using the nudged elastic band (NEB) method\cite{Henkelman_JCP001} to provide further evidence of a qualitative connection between the perturbation of the interatomic force constants and a flat area in the potential energy surface caused by adjacent minima.

The paper is arranged as follows: In Section II, we provide methodological and computational details. In section III, we present and analyze the results of our calculations. Finally, we extract the main conclusions in section IV.

\section{Methodology}
\subsection{Lattice thermal conductivity}
The lattice thermal conductivity tensor ($\kappa _{l}^{\alpha \beta}$) can be calculated in the relaxation time approximation from the expression
\begin{eqnarray}
\kappa _{l}^{\alpha \beta}=\frac{1}{k_B T^2 V}\sum_{j\vec{q}}n_0(n_0+1)(\hbar \omega_{j\vec{q}})^2 v^\alpha_{j\vec{q}} v^\beta_{j\vec{q}} \tau_{j\vec{q}},
\end{eqnarray}
where $V$ is the volume of the unit cell, $k_B$ is the Boltzmann constant, T is the temperature, $\alpha$ and $\beta$ denote Cartesian axes, $n_{0}$ is the Bose-Einstein occupancy, and $\omega$ and $v$ are an angular frequency and a group velocity respectively. The subscript $j$ runs over phonon branch indices and $\vec{q}$ runs over phonon wave vectors. The total phonon scattering rate $\tau_{j\vec{q}}^{-1}$ can be expressed as the sum of the contributions from different scattering mechanisms using Matthiessen's rule
\begin{eqnarray}
\label{Eq-sc}
\frac{1}{\tau_{j\vec{q}}}= \frac{1}{\tau^{\mathrm{anh}}_{j\vec{q}}} + \frac{1}{\tau^{\mathrm{iso}}_{j\vec{q}}} + \frac{1}{\tau^{\mathrm{def}}_{j\vec{q}}},
\end{eqnarray}
where $\tau^{\mathrm{anh}}_{j\vec{q}}$ has its roots in the intrinsic anharmonicity of the crystal, which enables three-phonon processes, $\tau^{\mathrm{iso}}_{j\vec{q}}$ is introduced by isotopic mass disorder and $\tau^{\mathrm{def}}_{j\vec{q}}$ is due to other defects present in the crystal. The three-phonon scattering rates can be calculated from the third-order interatomic force constants (IFCs) of the unperturbed crystal.\cite{almaBTE} The contribution due to isotopic mass disorder can be computed using the method developed by Tamura \textit{et al.}.\cite{TamuraPRB84} The calculation of the scattering rates due to the defects in the crystal is described in the next subsection.

\subsection{Phonon scattering by defects}
The phonon scattering rates due to defects can be obtained by employing the optical theorem from perturbation theory\cite{MingoPRB10}
\begin{eqnarray}
\label{Eq-simple}
\frac{1}{\tau^{\mathrm{def}}_{j\vec{q}}}=-\rho_{\mathrm{def}}V\frac{1}{\omega_{j\vec{q}}}\Im\{\braket{j\vec{q}|\mathbf{T}|j\vec{q}}\}
\end{eqnarray}
where $\rho_{\mathrm{def}}$ is the volume concentration of the point defect and $\omega$ the angular frequency of incident phonons. The $\mathbf{T}$ matrix connects the phonon wave functions of the pristine and perturbed systems, and can be calculated as
\begin{eqnarray}
\mathbf{T} = (\mathbf{I}-\mathbf{Vg^{+}})^{-1}\mathbf{V}.
\end{eqnarray}
Here, $\mathbf{g}^+$ is the retarded Green's function of the pristine system, and $\mathbf{V}$ is defined as
\begin{eqnarray}
\mathbf{V} = \mathbf{V}_M+\mathbf{V}_K.
\end{eqnarray}
$\mathbf{V}_M$ and $\mathbf{V}_k$ account for changes in mass and force constants between the perfect and defect-laden structures, respectively. $\mathbf{V}_M$ is diagonal, with elements
\begin{equation}
\mathbf{V}_{M,a} = -\frac{M_a^{\prime}-M_a}{M_a}\omega^2    
\end{equation}
and nonzero only for the on-site terms. $M^{\prime}$ and $M$ are the masses of the defect and of the original atom at the $a$-th site, respectively. The elements of the force-constant perturbation matrix are
\begin{equation}
\mathbf{V}_{K,ab}^{\alpha\beta} = \frac{K^{\prime\alpha\beta}_{a b}-K_{a b}^{\alpha\beta}}{\sqrt{M_a M_b}}
\end{equation}
 where $K^{\prime}$ and $K$ are the IFCs of the defect-laden and perfect structure respectively, and $a$, $b$ are atom indices. The 
$\mathbf{T}$-matrix can also be used to obtain the retarded Green's function of the perturbed system, $\mathbf{G}^+$, via the Dyson equation:
\begin{equation}
    \mathbf{G}^+=\mathbf{g}^+(\mathbf{I}+\mathbf{T}\mathbf{g}^+).
    \label{eq:Dyson}
\end{equation}

\subsection{Computational Details}
All the structural information and IFCs are extracted from density functional theory (DFT) calculations carried out using the projector-augmented plane wave method\cite{PAW94} with an energy cutoff of \SI{500}{eV} and the local density approximation to exchange and correlation as implemented in VASP.\cite{VASP196, VASP299} We consider both $4\times 4 \times 4$ and $5\times 5 \times 5$ supercells (containing 128 and 250 atoms, respectively) for the calculations of IFCs of the defect structures. The \textit{DX} centers are stable or metastable only in a negatively charged acceptor state,\cite{ParkPRB96} so we assume the $-1$ charge state for all the structures. The NEB calculation is only carried out using a $4\times 4 \times 4$ supercell.\cite{Henkelman_JCP001} For optimization, the total energy and force convergence criteria are set to \SI{1e-5}{eV} and \SI{1e-3}{eV.\angstrom^{-1}}. The calculated lattice parameter for GaAs is \SI{5.626}{\angstrom}, which matches well with other theoretical and experimental results.\cite{PierronAC66,MullinJAP76,NielsenPRB85} The supercell volume is kept fixed during the optimization of the defect structures. 

The second- and third-order IFCs are calculated through the direct method as implemented in the Phonopy \cite{phonopy} and thirdorder.py\cite{ShengBTE} codes respectively, with atomic displacements of \SI{0.01}{\angstrom}. For third-order IFCs, we consider up to 6th-nearest-neighbor interactions. For the non-analytical correction to the dynamical matrix, the Born effective charges and the dielectric tensor are calculated perturbatively with VASP.

After a careful convergence test, we settle on a grid of $33\times 33 \times 33$ q-points to compute the Green's function using the tetrahedron method,\cite{tetrahedron} and on a $29\times 29 \times 29$ grid to sample the incident phonons. Using these grids the calculated thermal conductivity was converged down to 40K The calculated defect scattering rates are used to obtain the final thermal conductivity using the almaBTE code.\cite{almaBTE} 

\section{Results and Discussion}
\begin{figure}[t]
\centerline{\hfill
    \includegraphics[width=0.490\textwidth]{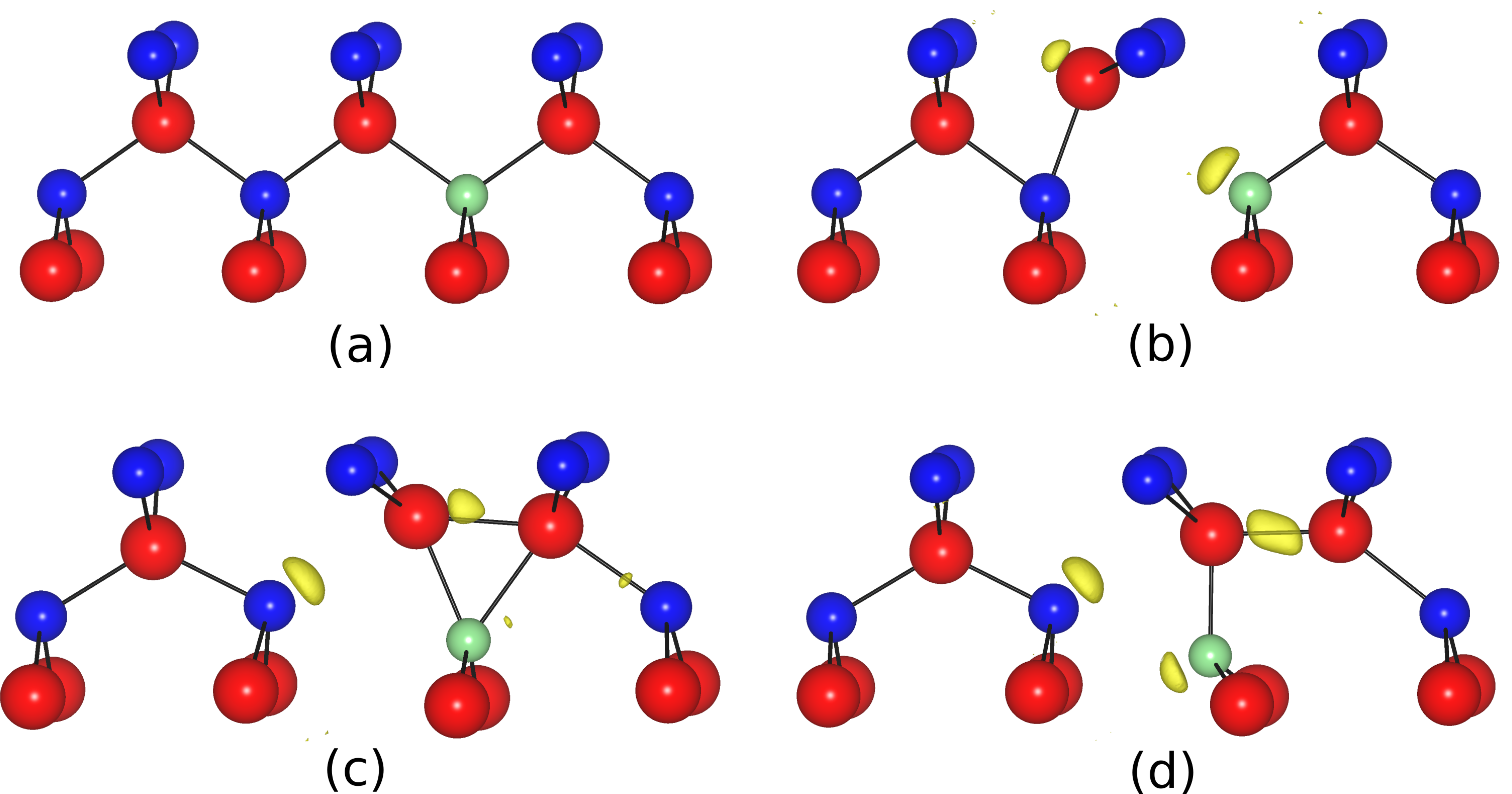}\hfill}
\caption{Atomic configurations and electron localization function (ELF) of the (a) $T_d$, (b) BB-\textit{DX}, (c) $\alpha$-CCB-\textit{DX} and (d) $\beta$-CCB-\textit{DX} structures in S-doped GaAs. Red, blue and green atoms represent Ga, As and S respectively. The ELF isosurfaces are set at a value of $0.90$.}
\label{struct}
\end{figure}

We calculate the lattice thermal conductivity in S-, Se- and Te- doped GaAs. We consider four different defect structures in each case. We study the $T_d$ defect, where the impurity atom simply replaces the As atom in the zincblende structure, and the more complex broken-bond (BB) and  $\alpha$- and $\beta$-cation-cation-bond (CCB) \textit{DX}-center structures. All the structures are fully relaxed, and the resulting configurations for GaAs:S are shown in Fig.~\ref{struct}. The optimized atomic structures of the BB-\textit{DX} and $\beta$-CCB-\textit{DX} defects [Fig.~\ref{struct}(b and d)] closely reproduce those in previous reports.\cite{ParkPRB96,DuPRB05,MaPRB13} The $\alpha$-CCB-\textit{DX} case in Fig.~\ref{struct}(c) is slightly different from the previously reported structure\cite{ParkPRB96,DuPRB05} in that we find an asymmetric relaxation. The energy of our asymmetric structure is only \SI{0.3}{meV} lower than the previously reported structure. However, the asymmetric relaxation is necessary for the $\alpha$-CCB-\textit{DX} center to be mechanically stable and to obtain purely real phonon frequencies.

The ELFs\cite{ELF} for all the \textit{DX} centers are also shown in Fig.~\ref{struct}. The ELF yields an estimate of the local effect of Pauli repulsion on the behavior of the electrons and allows a real-space mapping of the core, bonding and non-bonding regions in a crystal as well as an understanding of the nature of bonding and the presence of lone pairs of electrons.\cite{Savin_Angew97} The formation of electronic lone pairs plays an important role in the stabilization of the individual \textit{DX} center structures.\cite{Onopko01} Fig.~\ref{struct}(b) illustrates how, in the BB-\textit{DX} case, the bond between the dopant and Ga atom breaks due to the formation of lone pairs on the dopant and the neighboring Ga atom and how the repulsion  from the negatively charged sulfur defect causes the neighboring Ga to occupy an interstitial position. For the $\alpha$- and $\beta$-CCB-\textit{DX} centers, a similar picture of bond breaking is observed, as shown in Fig.~\ref{struct}(c) and (d). In this case, lone pairs are formed on the dopant and on an As atom. Moreover, the ELF reflects a strong cation-cation (Ga-Ga) covalent bond. Thus, complex lattice relaxations around the dopant are involved in the \textit{DX} centers which, as mentioned above, are believed to have a strong effect on thermal conductivity.

\begin{figure}[t]
\includegraphics[width=0.43\textwidth]{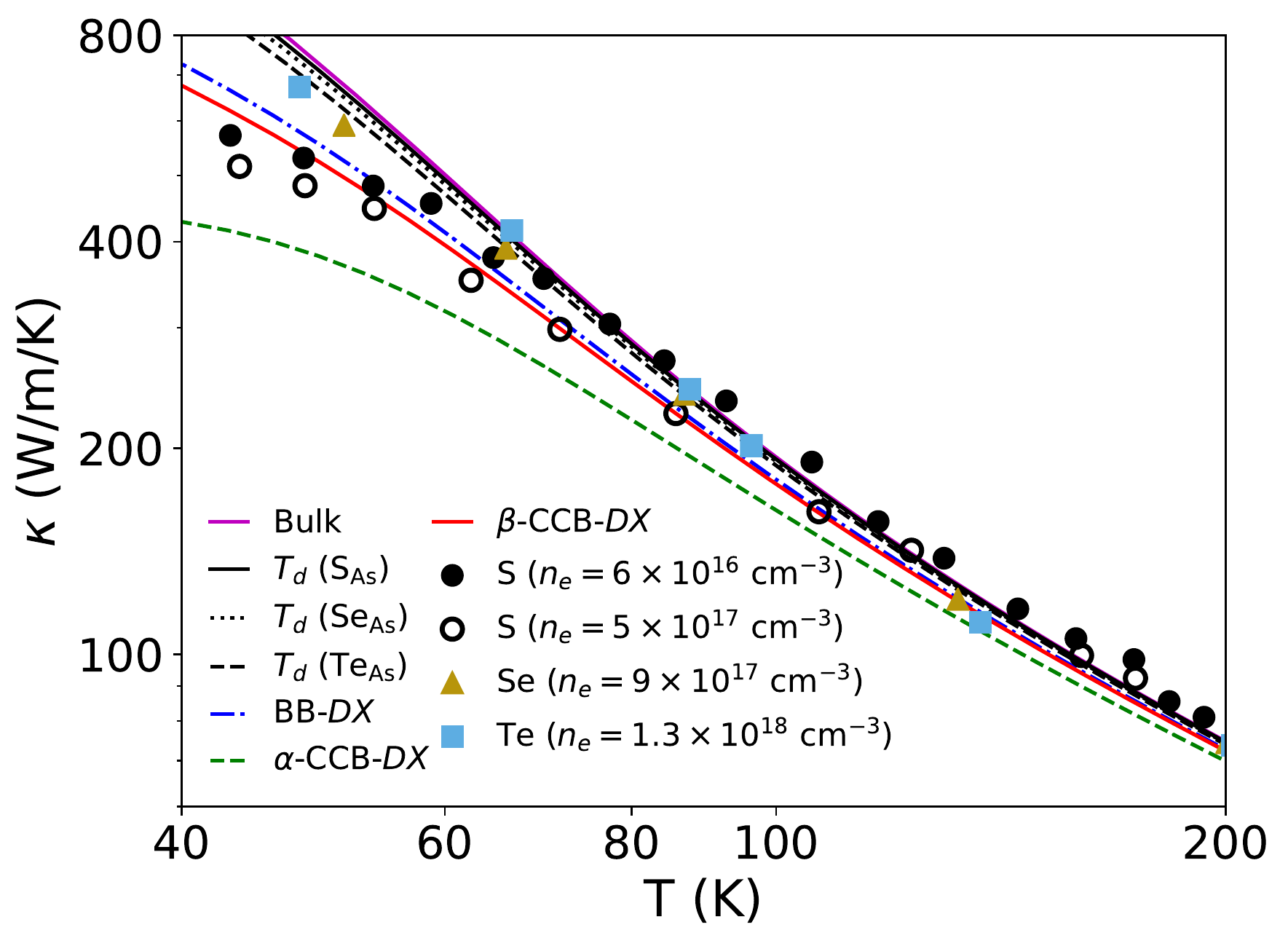}
\caption{Variation of the calculated thermal conductivity with temperature for pristine bulk GaAs, for all the different structures of GaAs:S and for the $T_d$ version of GaAs:Se and GaAs:Te, compared to experiment when available. The experimental data for GaAs:S are obtained from Ref.~\onlinecite{VuillermozPSSA77} and those for GaAs:Se and GaAs:Te come from Ref.~\onlinecite{CarlsonJAP65}. For GaAs:S, the experimental carrier concentrations are \SI{6e16}{cm^{-3}} (filled circles) and \SI{5e17}{cm^{-3}} (empty circles). The theoretical curves for the GaAs:S $T_d$ and \textit{DX} centers assume concentrations of \SI{5e17}{cm^{-3}}.}
\label{expt-tc}
\end{figure}

The thermal conductivity is calculated for all four defect structures for S-, Se- and Te-doped GaAs. The results for the $T_d$ defect in S-, Se- and Te-doped GaAs and for the \textit{DX} centers of GaAs:S are shown in Fig.~\ref{expt-tc} and compared with the experimental values obtained from single crystal samples.\cite{VuillermozPSSA77,CarlsonJAP65} The calculated thermal conductivity of pristine bulk GaAs is also plotted as a reference. In the region above \SI{100}{K}, where the thermal conductivity is dominated by three-phonon processes, the thermal conductivities for the defect-laden structures are very close to each other and to the bulk values, and are inversely proportional to the temperature. Theoretical calculations are performed at defect concentrations corresponding to the experimentally reported carrier concentrations\cite{VuillermozPSSA77,CarlsonJAP65} and it can be seen that the thermal conductivities of bulk and $T_d$-defect-containing GaAs are almost identical over the whole range of temperature. Even in the low-temperature region (\SIrange{40}{80}{K}) the influence of the $T_d$ defects is small. The experimental values for Se- and Te-doped samples\cite{CarlsonJAP65} only show signs of weak defect scattering and match well with our calculated values for the simple $T_d$ substitutional defect. On the other hand, the calculated thermal conductivity for different \textit{DX} centers in GaAs:S deviates significantly from the bulk thermal conductivity in the low-temperature region. Fig.~\ref{expt-tc} shows that a similar suppression of the low-temperature lattice thermal conductivity is observed experimentally for GaAs:S.\cite{VuillermozPSSA77} This clearly points to the \textit{DX} centers having very different strengths as phonon scatterers.  It is interesting to note that a similar depression of the experimental thermal conductivity in HgSe in the temperature range from \SIrange{30}{45}{K} was attributed to a resonance in the phonon-defect scattering rates.\cite{Nelson_PRL69} We have also considered boundary scattering as a possible alternative explanation of the low temperature thermal conductivity. However, to obtain a satisfactory agreement with experiment (see supplementary material), average grain sizes below 1~mm were necessary, which seems unreasonable as the experiments were performed on high quality single crystals with reported sizes around 25~mm.\cite{VuillermozPSSA77}

\begin{figure}[t]
\centerline{\hfill
    \includegraphics[width=0.49\textwidth]{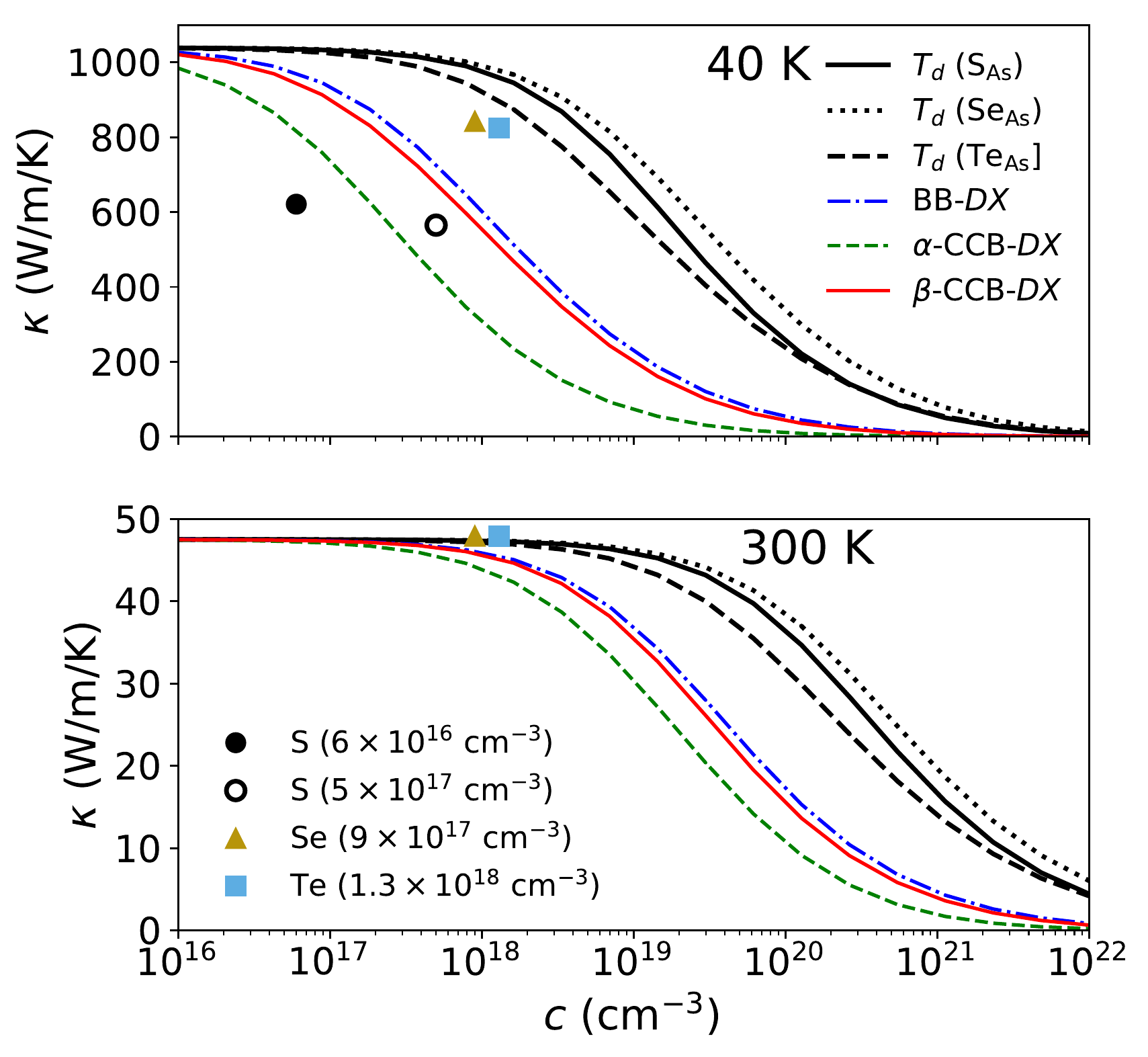}\hfill}
\caption{Variation of thermal conductivity with defect concentration for all the $T_d$ defects considered and the \textit{DX} centers formed by GaAs:S. The thermal conductivity is calculated at \SI{40}{K} and \SI{300}{K}. The experimental point are plotted at their measured carrier concentration.}
\label{tc}
\end{figure}
In the experimental reports used as sources for the data in Fig.~\ref{expt-tc} there is no information about the type of defect present in the sample. While the carrier concentration obtained from Hall coefficient measurements gives an indication of the defect concentration it may not directly correspond to the actual values. This will be the case when \textit{DX} centers and substitutional defects, which act as acceptor and donors respectively, compensate each other. We have previously proposed a compensation scenario to explain the thermal conductivity of FeSi.\cite{Stern_PRB18} In Fig.~\ref{tc} we show the variation of thermal conductivity as a function of carrier concentration for all the $T_d$ defects considered and for the \textit{DX} centers of GaAs:S at two different temperatures, \SI{40}{K}, and \SI{300}{K}.
Among the $T_d$ defects in Fig.~\ref{tc}, the Te-doped one leads to the largest reduction of thermal conductivity, followed by S and Se. This can be understood simply in terms of the mass difference between As and the doped atom, which takes values of \SI{43}{u}, \SI{4}{u} and \SI{53}{u} for the S, Se, and Te atoms, respectively. The thermal conductivity is substantially lowered by BB-\textit{DX}, $\beta$-CCB-\textit{DX} and $\alpha$-CCB-\textit{DX} GaAs:S centers, in that order. At the low carrier concentrations found in experimental measurements, the difference is only observable at low temperatures where phonon scattering by \textit{DX} centers dominates. It can be seen that the experimental data for GaAs:S ($\kappa \approx 600$~W/mK at 40~K) can be explained by the presence of \textit{DX}-centers at concentrations between $3\times 10^{17}$~cm$^{-3}$ and $1\times 10^{18}$~cm$^{-3}$. This would fit very well into a picture where the Fermi level is pinned to the center of the band gap by compensating and oppositely charged defects. This would lead to both a low defect concentration, due to the 1.5~eV band gap of GaAs, and to an even lower carrier concentration, due to the compensation between $T_d$-donor and \textit{DX}-acceptor defects. On the other hand, if the thermal conductivity were to be explained by substitutional $T_d$-defects alone, a defect concentration close to 10$^{20}$~cm$^{-3}$ would be necessary, Fig.~\ref{tc}. It is thus clear that a picture of uncompensated substitutional defects cannot simultaneously explain the low carrier concentration and the reduced thermal conductivity.
Similar behaviors are observed in the cases of Se and Te, and the corresponding figures are shown in the supplementary material. More specifically, the lowest thermal conductivities for a given concentration are obtained for the $\alpha$-CCB-\textit{DX} and $\beta$-CCB-\textit{DX} centers in Se and Te, respectively.

\begin{figure}[t]
\centerline{\hfill
    \includegraphics[width=0.490\textwidth]{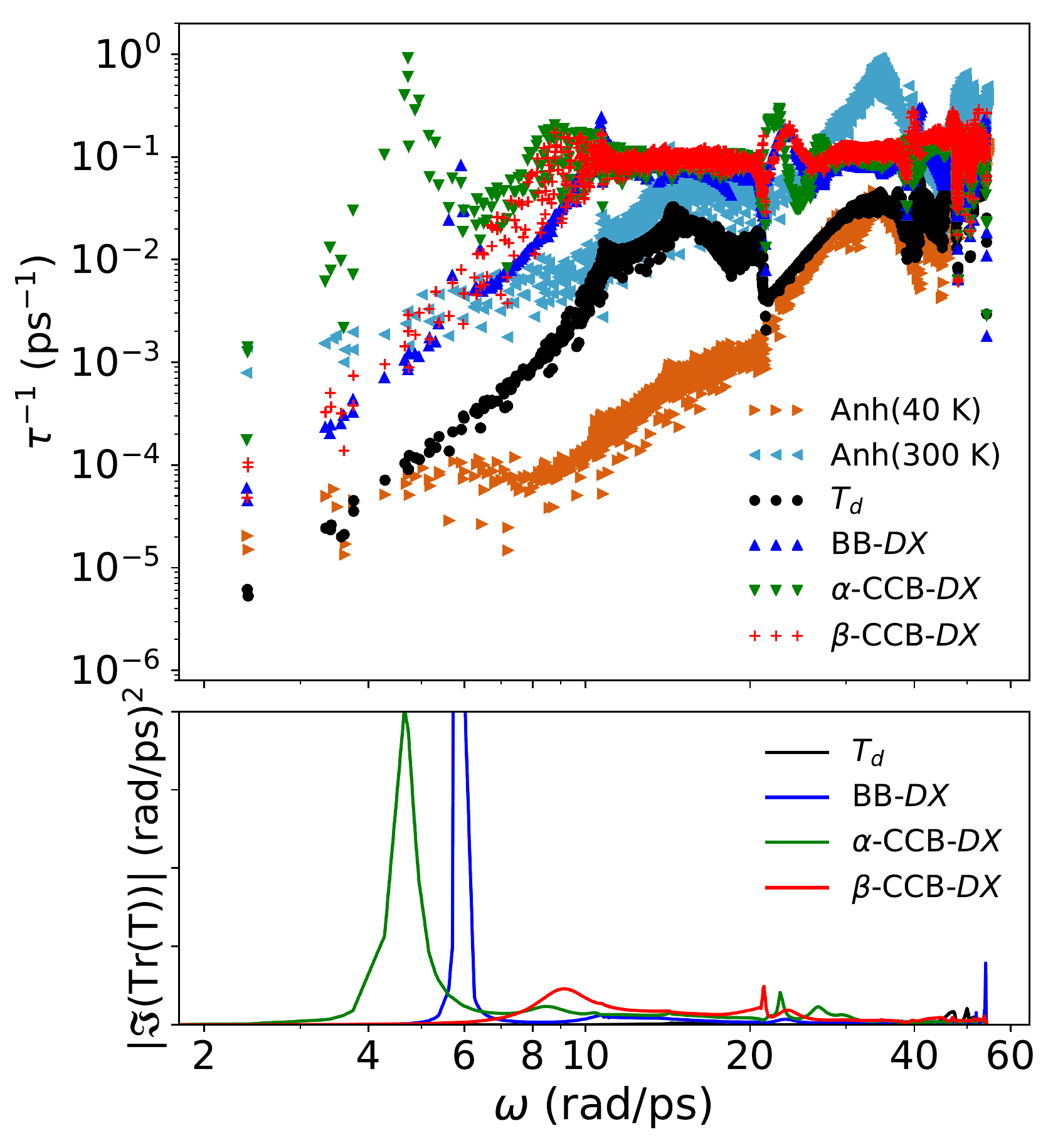}\hfill}
\caption{Phonon scattering rates and trace of the imaginary part of the $\mathbf{T}$ matrix for the $T_d$ defect and the \textit{DX} centers of GaAs:S as a function of phonon angular frequency. Defect concentrations of \SI{1e20}{cm^{-3}} are assumed. This defect concentration is only chosen for graphical clarity. Changing the defect concentration would
rigidly shift the scattering rates by the same magnitude. The phonon scattering rates introduced by those defects are also compared with the anharmonic scattering rates computed at \SI{40}{K} and \SI{300}{K}. }
\label{scr}
\end{figure}
A lowered thermal conductivity points to more intense phonon scattering. Fig.~\ref{scr} shows the scattering rates due to defects at a hypothetical concentration of $\SI{1e20}{cm^{-3}}$ and the trace of the imaginary part of the $\mathbf{T}$ matrix for the $T_d$ and \textit{DX} centers of GaAs:S. The anharmonic scattering rates at \SI{40}{K} and \SI{300}{K} are shown in order to illustrate the dominance of different scattering mechanisms at different temperatures. Most noticeably, the scattering rates caused by the $\alpha$-CCB-\textit{DX} exhibit a prominent peak at a frequency of \SI{4.7}{rad.ps^{-1}}. This peak is the signature of resonant phonon scattering and perfectly matches the corresponding peak in the elements of the $\mathbf{T}$ matrix. In fact, a marked peak in the imaginary part of the trace of the scattering \textbf{T} matrix has been shown to be a straightforward way to identify the resonance scattering.\cite{Katre_PRL17,Dongre_JMCC18} A very sharp peak also appears in the BB-\textit{DX} case at about \SI{5.9}{rad.ps^{-1}}, whereas the anomalously high scattering rates of $\beta$-CCB-\textit{DX} are mainly found around 8~rad/ps, close to where the transverse acoustic band of GaAs enters the edge of the Brillouin zone.\cite{Arrigoni_PRB18} As can be seen by comparing Figs.~\ref{tc} and \ref{scr}, the order of magnitude of the scattering rates in the low-frequency range has a significant impact on the thermal conductivity. It is clear that all the \textit{DX}-centers have a higher scattering rate, and result in a lower thermal conductivity, than the substitutional $T_d$ defect. It is also clear that the resonance for the BB-\textit{DX} falls in such a narrow frequency range that it affects the thermal conductivity much less than the large broad peak found in case of the $\alpha$-CCB-\textit{DX}. 
In the case of Se, we observe a similar trend in the thermal conductivity, likewise explained by the magnitude of the scattering rates in the low-frequency region (Figs.~2 and 3, supplementary material). For Te, we observe a different trend, where the lowest value of the thermal conductivity is registered for the $\beta$-CCB-\textit{DX} center (Fig.~2, supplementary material). However, this can also be understood from the scattering rates (Fig.~3, supplementary material) since $\beta$-CCB-\textit{DX} introduces scattering rates that are two orders of magnitude higher than those of BB-\textit{DX} and $\alpha$-CCB-\textit{DX} for the same concentration. In the following we will look into the origin of the narrow peak.

\begin{figure}[t]
\centerline{\hfill
    \includegraphics[width=0.49\textwidth]{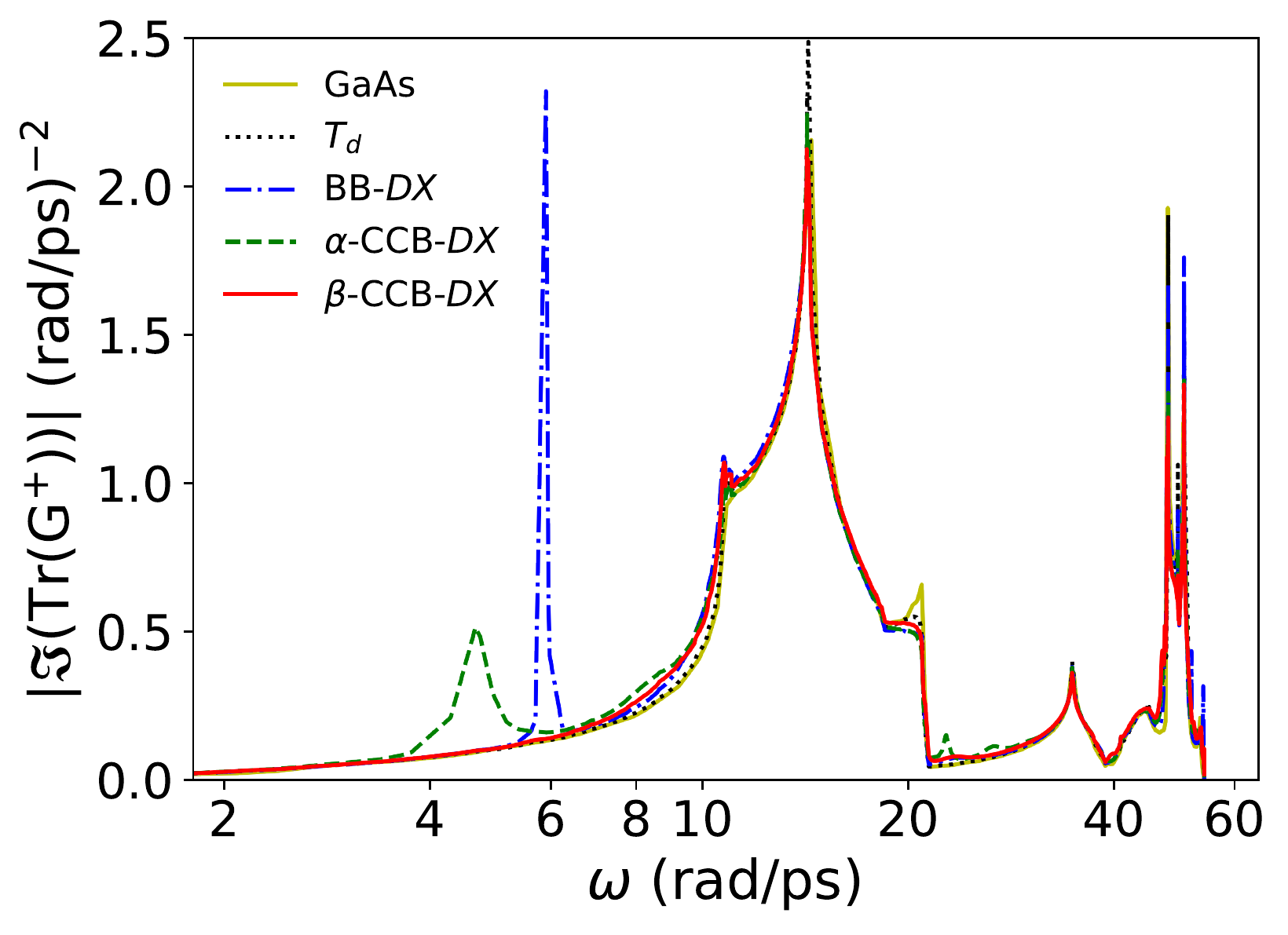}\hfill}
\caption{Trace of the imaginary part of the perturbed Green's function(\textbf{G}$^+$) for the $T_d$ defect and the \textit{DX} centers in GaAs:S. The phonon DOS of bulk GaAs is also shown for comparison.}
\label{green}
\end{figure}

\begin{figure}[t]
\centerline{\hfill
    \includegraphics[width=0.38\textwidth]{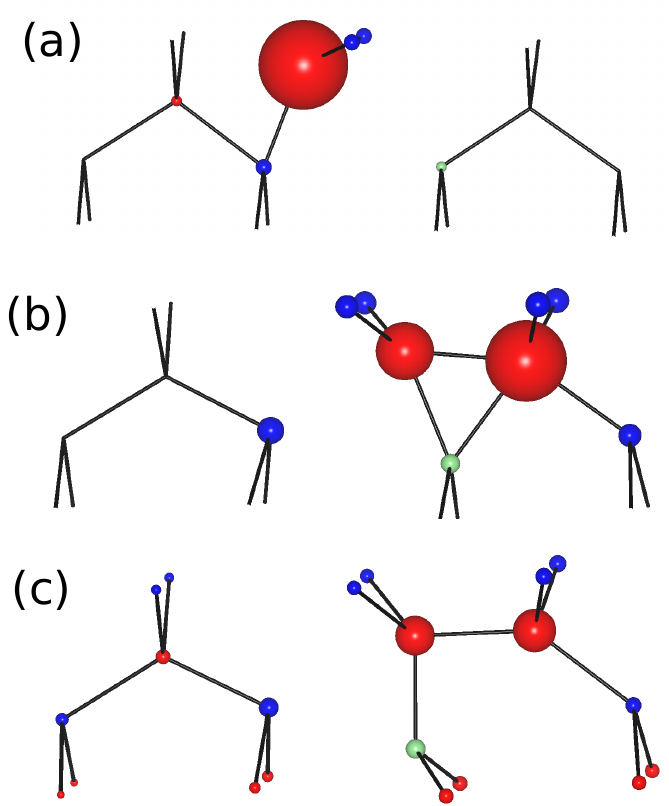}\hfill}
\caption{A schematic presentation of the perturbed projected densities of states on the atoms for (a) BB-\textit{DX}, (b) $\alpha$-CCB-\textit{DX} and (c) $\beta$-CCB-\textit{DX} in GaAs:S at a frequency corresponding to the peak in the frequency region from \SIrange{4}{10}{rad.ps^{-1}} (see Fig.~\ref{green}). The volume of the sphere representing each atom is proportional to the contribution of that atom to the peak. The color code is the same as in Fig.~\ref{struct}}
\label{proj_gf}
\end{figure}


To dig further into the nature of the resonances, we have calculated the Green's function of the perturbed system, ${\bf G}^+$, via the Dyson equation, Eq.~\eqref{eq:Dyson}. The trace of the imaginary part of the projection of ${\bf G}^+$ on the degrees of freedom of the $5\times 5\times 5$ supercell is shown in Fig.~\ref{green} as representative of the phonon density of states (DOS) of the perturbed system. Besides the propagating modes inherited from the unperturbed structure, the plot shows a clear peak at each frequency where a resonance is observed in the scatterings rates. Such peaks hint at more localized vibrations as causes of those resonances. To confirm this inference, in Fig.~\ref{proj_gf} we plot the individual DOS projected on the atoms making up the defect at a frequency corresponding to the peak, with each atom represented by a sphere with a volume proportional to its relative contribution to the phonon DOS. It can be seen that the $\delta$-like peak for BB-\textit{DX} (see Figs.~\ref{scr} and \ref{proj_gf}) corresponds to modes almost entirely localized around the Ga atom that is displaced to an interstitial position. The broader peak in the case of $\alpha$-CCB-\textit{DX} is associated to an intermediate degree of localization made possible by the asymmetric structural relaxations around the dopant.
The interaction (i.e., scattering) of incident phonons with those intermediately localized modes strongly hinders thermal transport. Finally the broad peak observed for $\beta$-CCB-\textit{DX} is quite delocalized over the surroundings of the defect. 

\begin{figure}[hb]
\centerline{\hfill
    \includegraphics[width=0.43\textwidth]{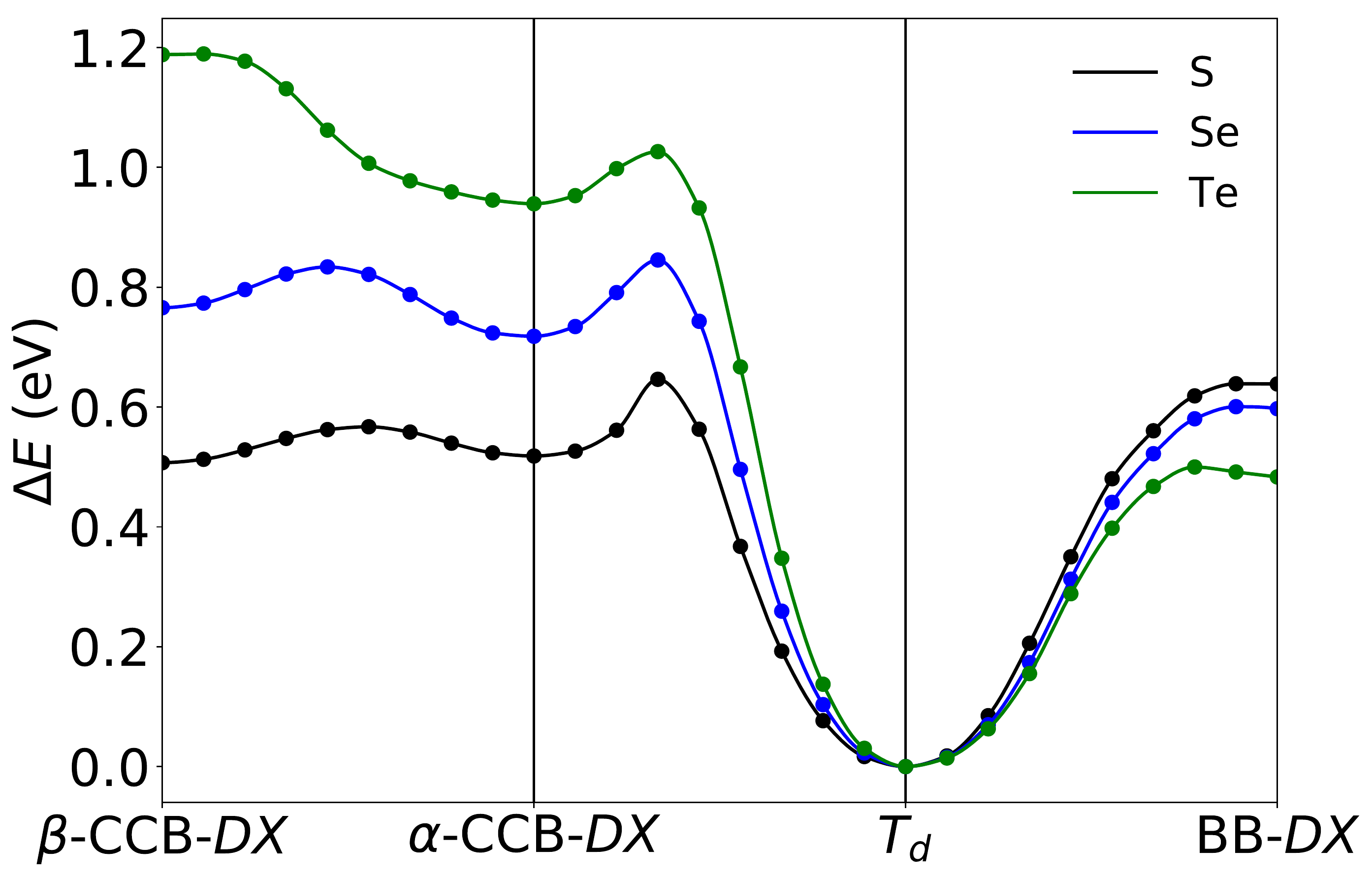}\hfill}
\caption{Nudged-elastic-band minimum energy paths connecting the defect structures formed by S, Se, and Te dopants in GaAs. $\Delta E$ is the energy of each configuration when the energy of the $T_d$ defect is taken as the baseline. The energies have not been corrected for the use of charged cells.\cite{Persson_PRB05} The correction due to the valence band maximum, $\Delta \varepsilon_{VBM}$, lowers the $\beta$-CCB, $\alpha$-CCB and BB-\textit{DX} sulphur defects by 1.21~eV, 0.95~eV and 0.35~eV respectively with respect to the $T_d$-defect. Thereby the $\beta$-CCB-\textit{DX} becomes the most stable defect for the $q=-1$ charged state.}
\label{enld}
\end{figure}
In a previous study, we posited that two or more close energy minima in the energy landscape are a necessary condition for having resonant phonon scattering since they create very flat-bottomed energy valleys.\cite{Dongre_JMCC18} Thus, we calculate the barriers to transitions between different configurations using the NEB method. Fig.~\ref{enld} shows the results for all structures and all dopants. It can be seen that the $T_d$ defect has the lowest energy among all the impurities and the \textit{DX} centers are thus only locally stable. We constructed intermediate structures between all four defects. However, we found that the paths connecting the CCB- and BB-\textit{DX} defects relax to the $T_d$ structure, which also seems reasonable considering the structures in Fig.~\ref{struct}. Still, the energy minima corresponding to the \textit{DX} centers are very flat. This is in line with our earlier analysis of the conditions necessary for finding an IFC perturbation large enough to produce resonances in the scattering rates in the low-frequency region for the \textit{DX} centers.\cite{Dongre_JMCC18}

\section{Conclusions}
 The formation of \textit{DX} centers in GaAs is explained by investigating the electron localization function, and we conclude that the lone pair electrons play a crucial role in determining their structure. We performed \textit{ab initio} calculations of the lattice thermal conductivity of GaAs in the presence of $T_d$ defects or \textit{DX} centers induced by S, Se, and Te dopants. The calculated thermal conductivity shows good agreement with existing measurements. 
 
 The asymmetric relaxation in the \textit{DX} centers causes strong perturbations in the IFCs resulting in intense phonon scattering even at low defect concentrations. The results thereby strengthen the emerging understanding of how the lattice thermal conductivity can be used to unveil the dominant defects in semiconductors. 
 
The resonances in the phonon scattering rates were attributed to localized vibrational modes in the perturbed system associated to the structural distortions. These localized modes are connected to flat valleys in the energy landscape such as those emerging from degenerate minima. Overall, we show the dramatic influence of the local atomic structure of the dopant on a macroscopic quantity like the thermal conductivity and illustrate how to take it into account,  which might be useful for advanced semiconductor design.

\section*{ACKNOWLEDGMENT}
A. K. thanks Bonny Dongre and Marco Arrigoni for their help in the calculations. The authors acknowledge support from the European Union’s Horizon 2020 Research and Innovation Action, Grant No. 645776 (ALMA), and from the French (ANR) and Austrian (FWF) Science Funds, project CODIS (ANR-17-CE08-0044-01 and FWF-I-3576-N36). We also thank the Vienna Scientific Cluster for providing the computational facilities (project number 70958: ALMA).

\bibliographystyle{aip} 


\begin{thebibliography}{10}

\bibitem{Sze_Book06}
S.~M. Sze and K.~K. Ng,
\newblock {\em Physics of semiconductor devices},
\newblock John wiley \& sons, 2006.

\bibitem{Brennan_Book99}
K.~F. Brennan,
\newblock {\em The physics of semiconductors: with applications to
  optoelectronic devices},
\newblock Cambridge university press, 1999.

\bibitem{Queisser_Science98}
H.~J. Queisser and E.~E. Haller,
\newblock Science {\bf 281}, 945 (1998).

\bibitem{Callaway_PRB60}
J.~Callaway and H.~C. von Baeyer,
\newblock Phys. Rev. {\bf 120}, 1149 (1960).

\bibitem{Katre_PRL17}
A.~Katre, J.~Carrete, B.~Dongre, G.~K.~H. Madsen, and N.~Mingo,
\newblock Phys. Rev. Lett. {\bf 119}, 075902 (2017).

\bibitem{Katre_PRM18}
A.~Katre, J.~Carrete, T.~Wang, G.~K.~H. Madsen, and N.~Mingo,
\newblock Phys. Rev. Materials {\bf 2}, 050602 (2018).

\bibitem{Polanco_PRB18}
C.~A. Polanco and L.~Lindsay,
\newblock Phys. Rev. B {\bf 98}, 014306 (2018).

\bibitem{Mooney_JAP90}
P.~M. Mooney,
\newblock J. Appl. Phys. {\bf 67}, R1 (1990).

\bibitem{CarlsonJAP65}
R.~O. Carlson, G.~A. Slack, and S.~J. Silverman,
\newblock J. Appl. Phys. {\bf 36}, 505 (1965).

\bibitem{VuillermozPSSA77}
P.~L. Vuillermoz, J.~Jouglar, A.~Laugier, and H.~R. Winteler,
\newblock Phys. Status Solidi (a) {\bf 41}, 561 (1977).

\bibitem{Dongre_JMCC18}
B.~Dongre, J.~Carrete, A.~Katre, N.~Mingo, and G.~K.~H. Madsen,
\newblock J. Mater. Chem. C {\bf 6}, 4691 (2018).

\bibitem{ParkPRB96}
C.~H. Park and D.~J. Chadi,
\newblock Phys. Rev. B {\bf 54}, R14246 (1996).

\bibitem{DuPRB05}
M.-H. Du and S.~B. Zhang,
\newblock Phys. Rev. B {\bf 72}, 075210 (2005).

\bibitem{MaPRB13}
J.~Ma and S.-H. Wei,
\newblock Phys. Rev. B {\bf 87}, 115210 (2013).

\bibitem{ELF}
A.~D. Becke and K.~E. Edgecombe,
\newblock J. Chem. Phys. {\bf 92}, 5397 (1990).

\bibitem{Henkelman_JCP001}
G.~Henkelman, B.~P. Uberuaga, and H.~J{\'o}nsson,
\newblock J. Chem. Phys. {\bf 113}, 9901 (2000).

\bibitem{almaBTE}
J.~Carrete, B.~Vermeersch, A.~Katre, A.~van Roekeghem, T.~Wang, G.~K.~H.
  Madsen, and N.~Mingo,
\newblock Computer Physics Communications {\bf 220}, 351  (2017).

\bibitem{TamuraPRB84}
S.-i. Tamura,
\newblock Phys. Rev. B {\bf 30}, 849 (1984).

\bibitem{MingoPRB10}
N.~Mingo, K.~Esfarjani, D.~A. Broido, and D.~A. Stewart,
\newblock Phys. Rev. B {\bf 81}, 045408 (2010).

\bibitem{PAW94}
P.~E. Bl\"ochl,
\newblock Phys. Rev. B {\bf 50}, 17953 (1994).

\bibitem{VASP196}
G.~Kresse and J.~Furthm\"uller,
\newblock Phys. Rev. B {\bf 54}, 11169 (1996).

\bibitem{VASP299}
G.~Kresse and D.~Joubert,
\newblock Phys. Rev. B {\bf 59}, 1758 (1999).

\bibitem{PierronAC66}
E.~D. Pierron, D.~L. Parker, and J.~B. McNeely,
\newblock Acta Crystallogr. {\bf 21}, 290 (1966).

\bibitem{MullinJAP76}
J.~B. Mullin, B.~W. Straughan, C.~M.~H. Driscoll, and A.~F.~W. Willoughby,
\newblock J. Appl. Phys. {\bf 47}, 2584 (1976).

\bibitem{NielsenPRB85}
O.~H. Nielsen and R.~M. Martin,
\newblock Phys. Rev. B {\bf 32}, 3792 (1985).

\bibitem{phonopy}
A.~Togo and I.~Tanaka,
\newblock Scr. Mater. {\bf 108}, 1 (2015).

\bibitem{ShengBTE}
W.~Li, J.~Carrete, N.~A. Katcho, and N.~Mingo,
\newblock Comp. Phys. Commun. {\bf 185}, 1747–1758 (2014).

\bibitem{tetrahedron}
P.~Lambin and J.~P. Vigneron,
\newblock Phys. Rev. B {\bf 29}, 3430 (1984).

\bibitem{Savin_Angew97}
A.~Savin, R.~Nesper, S.~Wengert, and T.~F. F{\"a}ssler,
\newblock Angew. Chemie Int. Ed. {\bf 36}, 1808 (1997).

\bibitem{Onopko01}
D.~Onopko and A.~Ryskin,
\newblock Semiconductors {\bf 35}, 1223 (2001).

\bibitem{Nelson_PRL69}
D.~A. Nelson, J.~G. Broerman, E.~C. Paxhia, and C.~R. Whitsett,
\newblock Phys. Rev. Lett. {\bf 22}, 884 (1969).

\bibitem{Stern_PRB18}
R.~Stern, T.~Wang, J.~Carrete, N.~Mingo, and G.~K.~H. Madsen,
\newblock Phys. Rev. B {\bf 97}, 195201 (2018).

\bibitem{Arrigoni_PRB18}
M.~Arrigoni, J.~Carrete, N.~Mingo, and G.~K.~H. Madsen,
\newblock Phys. Rev. B {\bf 98}, 115205 (2018).

\bibitem{Persson_PRB05}
C.~Persson, Y.-J. Zhao, S.~Lany, and A.~Zunger,
\newblock Phys. Rev. B {\bf 72}, 035211 (2005).

\end{thebibliography}

\end{document}